\documentclass[11pt,a4paper,twoside]{article}
\usepackage{graphicx}
\usepackage{color}
\usepackage{epstopdf}
\usepackage{lscape}
\usepackage{amsmath}
\usepackage{hyperref}
\usepackage{setspace}
\begin{document}
\baselineskip=0.20in
\vspace{20mm}
\baselineskip=0.30in
\begin{center}
{\large \bf Exact solution of the Schr\"{o}dinger equation with the $q$-deformed quantum potentials using the Nikiforov--Uvarov method}
\vspace{4mm}

{ \large {\bf Falaye, B.\ J.}\footnote{E-mail:~ fbjames11@physicist.net} , {\bf Oyewumi, K.\,J.}\footnote{E-Mail:~ kjoyewumi66444@unilorin.edu.ng} and {\bf Abbas, M.}\footnote{E-Mail:~ abbasmustapha1@yahoo.com}} \vspace{5mm}

{Theoretical Physics Section, Department of Physics\\ University of Ilorin,  P. M. B. 1515, Ilorin, Nigeria }

\vspace{4mm}

\end{center}

\noindent
\begin{abstract}
\noindent
In this paper, we present the exact solution of the one-dimensional Schr\"{o}dinger equation for the $q$-deformed quantum potentials via the Nikiforov--Uvarov method. The eigenvalues and eigenfunctions of these potentials are obtained via this method. The energy equations and the corresponding wave functions for some special cases of these potentials are briefly discussed. The PT-symmetry and Hermiticity for these potentials are also discussed.
\end{abstract}


{\bf PACs No.} 03.65.Ge, 02.30.Gp

\section{Introduction}
It is well known that the exact solutions of the Schr\"{o}dinger equation for some physical potentials play an essential role in non-relativistic quantum mechanics and this is possible only for a few set of quantum mechanical systems. In recent years, a number of researchers have studied the solution of the Schr\"{o}dinger equations with various potentials for different applications. These investigations include: Woods--Saxon Potential [1 - 6], Harmonic oscillator potential \cite{Ahm02}, Hulth$\acute{e}$n potential [8 - 12], Rosen--Morse potential [13 - 19], Manning-Rosen potential [20 - 27], Hulth$\acute{e}$n plus Manning-Rosen \cite{MeD09}, Kratzer potential \cite{Oye10}, Pseudoharmonic potential \cite{OyE08, OyS12}, trigonometry Scarf potential \cite{FaO11}, Scarf-type potential \cite{MoA10}, Scarf--Grosche potential \cite{YaO12}, external hyperbolic-tangent potential \cite{WE1} and Yukawa potential \cite{N7, N8}.

Methods, such as the Nikivorov-Uvarov (NU) method \cite{BeE08, BeE06, BeE05, IkS10, Agb09, Agb11, AkM08, AkE10, Ikh10, IkS12, MeD09, IkA11, FaO11, MoA10, YaO12,NiU88, NiE91, N9, N10, N11, N12}, supersymetric approach \cite{B1, B3, B3, B4, B5, B6, B7, B8, B9,B10,B11,B12,B13,B14,B15,B16}, exact quantization rule \cite{B17,B18,B19,B20,B21,B22}, asymptotic iteration method \cite{B23,B24,B24,B25,B26,B27,B28,B29,B30,B31,C1,B32,B33,B34} and others have been some of the utility tools in solving the Schr\"{o}dinger equation, Klein-Gordon and Dirac equation with potential of interest. In this paper, we adopt the Nikiforov--Uvarov method, which has been used to solve several quantum mechanical problems in physics and applied mathematics.

By means of the Nikiforov--Uvarov method, the solutions of the Dirac equation with spin and pseudospin symmetries for the trigonometric Scarf potential in $D$-dimensions are obtained by Falaye and Oyewumi (2011) \cite{FaO11}. In addition, Ikhdair and Sever (2012) (and the references therein) used the NU method to obtain the approximate analytical solutions of the Dirac equations with the reflectionless-type and Rosen--Morse potentials including the spin--orbit centrifugal (pseudo-centrifugal) term \cite{IkS12}.

Very recently, Abdalla \emph{et al.} \cite{B35} derive analytical solutions of the Schr\"{o}dinger wave equation for some $q$-deformed potentials in terms of Heun functions. In this paper, our  solve the one-dimensional Schr\"{o}dinger equation with the $q$-deformed quantum potential (i.e., the Woods--Saxon plus Rosen--Morse plus symmetrical double well potential) using the Nikiforov--Uvarov method. In addition, by extension, other interesting related results  are obtained.

The rest of the paper is organized as follows: in Section~2, the NU method is reviewed. In Section~3, the bound state solutions are obtained for real, complex, PT-symmetric and non PT-symmetric cases of this potential. Section~4 contains discussion on some special cases. The conclusion is given in Section~ 5.

\section{NU Method}
The NU method is based on solving a second-order linear differential equation by reducing it to a generalized equation of the hypergeometric type \cite{NiU88, NiE91}. By introducing an appropriate coordinate transformation $z = z(r)$, this equation can be re-written in the following form:
\begin{equation}
\Psi''(z) + \frac{\tilde{\tau}(z)}{\sigma(z)} \Psi '(z) + \frac{\tilde{\sigma}(z)}{\sigma^{2}(z)}\Psi(z)=0
\label{E1},
\end{equation}	
where $\sigma(z)$ and $\tilde{\sigma}(z)$ are polynomials of at most of second degree, and $\tilde{\tau}(z)$  is a first degree polynomial. Taking the following factorization:	
\begin{equation}
\Psi(z) = y(z) \phi(z) ,
\label{E2}
\end{equation}	
equation (\ref{E1}) reduces to the equation of the hypergeometric type
\begin{equation}
\sigma(z) y''(z) + \tau(z)y'(z) + \lambda y(z) = 0
\label{E3},
\end{equation}
where
\begin{equation}
\tau(z) = \tilde{\tau}(z) + 2 \pi(z)
\label{E4}.
\end{equation}
The function $\pi$ and the parameter $\lambda$ in this method are defined as follows:
\begin{equation}
\pi(z) = \left(\frac{\sigma'(z) - \tilde{\tau}(z)}{2}\right) \pm  \sqrt{\left( \frac{\sigma'(z) - \tilde{\tau}(z)}{2}\right)^{2} + \tilde{\sigma}(z) + k \sigma(z) },
\label{E5}
\end{equation}
and
\begin{equation}
\lambda = k + \pi'(z)
\label{E6}.
\end{equation}

In order to find the value of $k$, the expression under the square root must be a square of a polynomial, this gives the polynomial $\pi(z)$  which is dependent on the transformation function $z(r)$. In addition, the parameter $\lambda$ defined in Eq.~ (\ref{E6}) takes the form
\begin{equation}
\lambda =\lambda_{n} = - n \tau' - \left[ \frac{n(n - 1)}{2} \sigma '' \right].
\label{E7}
\end{equation}
The polynomial solutions $y_{n}(z)$ are given by the Rodrigue relation
\begin{equation}
y_{n}(z) = \frac{N_{n}} {\rho(z)} \frac{d^{n}}{dz^{n}} \left[\sigma^{n}(z) \rho(z) \right]
\label{E8},
\end{equation}
where $N_{n}$ is the normalization constant and $\rho(z)$ is the weight function satisfying
\begin{equation}
\frac{d}{dz}\left[\sigma(z) \rho(z) \right] = \tau(z)\rho(z)
\label{E9}.
\end{equation}
The second part of the wave function $\phi(z)$  can be obtained from the relation
\begin{equation}
\pi(z) = \sigma(z) \frac{d}{dz}[\ln \phi(z)]
\label{E10}.
\end{equation}
This method has been a very useful tool in the fields of physics and applied mathematics since its introduction \cite{BeE08, BeE06, BeE05, IkS10, Agb09, Agb11, AkM08, AkE10, Ikh10, IkS12, MeD09, IkA11, FaO11, MoA10, YaO12,NiU88, NiE91}.
\section{Bound-state solutions}
\subsection{Exact solutions of the Schr\"{o}dinger equation with the $q$-deformed quantum potentials}
The one-dimensional time independent Schr\"{o}dinger equation  ($l = 0$) can be written as ($ \hbar = 2\mu = 1$)
\begin{equation}
\frac{d^{2} \Psi (x)}{dx^{2}} + [ E - V(x)]\Psi (x) = 0,
\label{E11}
\end{equation}
where $E$ is the energy, $V(x)$ is the potential energy, and $\psi(x)$ is the corresponding wave function. We consider the $q$-deformed quantum potentials, that is, the $q$-deformed generalized Woods--Saxon plus Rosen--Morse plus generalized symmetrical double well Potential which can be express as
\begin{equation}
V(x) =
\frac{V_{1}{\rm e}^{-2 \alpha x}}{1 + q {\rm e}^{-2\alpha x}} + \frac{V_{2} {\rm e}^{-4\alpha x}}{(1 + q{\rm e}^{-2\alpha x})^2} + V_{3} \sec h_{q}^2(\alpha x) + V_{4} \tan h_{q}(\alpha x) - V_{5} \tan h_{q}^2(\alpha x) + V_{6} \sec h_{q}^2(\alpha x)
\label{E12}.
\end{equation}
Here, the following deformed hyperbolic function described by Arai in (1991) has been used \cite{Ara91}:
\begin{eqnarray}
&\sin h_{q}{x} = \frac{{\rm e}^{x} - q{\rm e}^{-x}}{2}, ~\cos h_{q}{x} = \frac{{\rm e}^{x} + q {\rm e}^{-x}}{2},~\tan h_{q}{x} = \frac{\sin h_{q}{x}}{\cos h_{q}{x}}, ~\sec h_{q}{x} = \frac{1}{\cos h_{q}{x}},
\nonumber \\
&\mbox{cosec} h_{q}{x} = \frac{1}{\sin h_{q}{x}},~\cot h_{q}{x} = \frac{\cos h_{q}{x}}{\sin h_{q}{x}},~\cosh_{q}^{2}{x} - \sinh_{q}^{2}{x} = q,~ \frac{d}{dx}\cosh_{q}{x} = \sinh_{q}{x}, \nonumber \\
&~ \frac{d}{dx}\sinh_{q}{x} = \cosh_{q}{x},~\frac{d}{dx}\tanh_{q}{x} = \frac{q}{\cosh_{q}^{2}{x}}, ~\frac{d}{dx}\coth_{q}{x} = -\frac{q}{\sinh_{q}^{2}{x}}.
\label{E13}
\end{eqnarray}
By using the transformation
\begin{equation}
z = -{\rm e}^{-2\alpha x}
\label{E14},
\end{equation}
and substituting Eqs.~ (\ref{E14}) and (\ref{E12}) and appropriate parameters in Eq.~(\ref{E13}) into Eq.~(\ref{E11}), we have
\begin{eqnarray}
&\displaystyle{\Psi^{''}(z) + \frac{1 - qz}{z(1 - qz)} \Psi^{'}(z) + \frac{1}{z^{2}(1 - qz)^{2}}\left[z^{2}\left\{ \beta _{1} q - \beta _{2} - q^{2}(\beta _{4} + \beta _{5} + \xi) \right\}\right]} \nonumber \\
&  \displaystyle {+ \frac{1}{z^{2}(1 - qz)^{2}}\left[z \left\{ -\beta _{1} - 4(\beta_{3} + \beta _{6}) - 2q \beta _{5} + 2q \xi  \right\} + (\beta _{4} - \xi)\right] = 0}.
\label{E15}
\end{eqnarray}
where
\begin{equation}
\xi =\frac{-E}{4 \alpha^{2}}~\mbox{and}~ \beta_{i} = \frac{V_{i}}{4 \alpha^{2}}, ~ i = 1,~2,~3,~4,~5,~6.
\label{E16}
\end{equation}
Comparing Eq.~(\ref{E16}) with Eq.~(\ref{E1}), we have
\begin{eqnarray}
&\sigma(z) = z^{2}\left\{ \beta _{1} q - \beta _{2} - q^{2}(\beta _{4} + \beta _{5} + \xi) \right\}
+ z \left\{ -\beta _{1} - 4(\beta_{3} + \beta _{6}) - 2q \beta _{5} + 2q \xi  \right\} + (\beta _{4} - \xi), \nonumber \\ &\tilde{\tau}(z)=1-qz,~ \sigma(z) = z - qz^{2}
\label{E17}.
\end{eqnarray}
Inserting the values of these parameters with $\sigma'(z)=1-2qz$ into Eq.~(\ref{E5}), we have
\begin{equation}
\pi(z) = -\frac{qz}{2} \pm \frac{1}{2}b\left[ \{ 2\delta - \eta H \}qz - 2\delta \right],~ \mbox{for}~k = 2q (\beta _{4} - \beta_{5}) - 4(\beta_{3} + \beta_{6}) - \beta_{1} + \eta Hq \delta
\label{E18},
\end{equation}
where
\begin{equation}
\eta = \pm 1,~ \delta = \sqrt{\xi - \beta_{4}},~H = \sqrt{\frac{4 \beta_{2}}{q^{2}} + 16\frac{(\beta_{3} + \beta_{6} )}{q} + 12\beta_{5} + 1 }
\label{E19}~.
\end{equation}
Selecting an appropriate parameter of $k$ in $\pi(z)$ which satisfies the condition $\tilde{\tau}(z)<0$, one obtains
\begin{equation}
\pi (z) = -\frac{qz}{2} - \frac{1}{2} \left[ \{\delta - \eta H \}qz - 2\delta \right]~ \mbox{for}~k = 2q ( \beta _{4} - \beta_{5}) - 4 ( \beta_{3} + \beta_{6}) - \beta_{1} + \eta Hq\delta.
\label{E20}
\end{equation}
From Eq.~(\ref{E6}), parameter $\lambda$ is obtained as
\begin{equation}
\lambda = 2q(\beta_{4} - \beta_{5}) - 4(\beta_{3} + \beta_{6}) - \beta_{1} + \eta Hq\delta - \frac{q}{2} - \frac{1}{2} \left\{ \{ 2\delta - \eta H \}qz \right\},
\label{E21}
\end{equation}
and using Eq.~(\ref{E7}), we have
\begin{equation}
\lambda = \lambda_{n} = nq\left\{ (n + 1) + 2\delta - \eta H \right\}
\label{E22}.
\end{equation}
By comparing Eqs.~(\ref{E21}) and (\ref{E22}), the exact energy eigenvalue equation is obtained as
\begin{eqnarray}
&&E_{n} = -\frac{\alpha^{2}}{4}\left[ \frac{\frac{2qV_{4} + qV_{5} - V_{1}}{q \alpha ^{2}} + \frac{V_{2}}{q^{2} \alpha ^{2}}}{2n + 1 - \eta \sqrt{H}}-\left(2n + 1 - \eta \sqrt{H} \right) \right]^{2}-V_{4};\ \ \ n\ge 0,~q\ge 1\nonumber
\label{E23},
\end{eqnarray}
where we have substituted for $\beta _{1},~ \beta _{2},~\beta _{3},~ \beta_{4},~ \beta_{5},~ \beta_{6},~ \delta $ and
$H$. From Eq.~(\ref{E9}), it can be shown that the weight function $\rho(z)$ is
\begin{equation}
\rho(z) = z^{2\delta}{(1 - qz)}^{-\eta H}
\label{E24}.
\end{equation}
Substituting Eq.~(\ref{E24}) into the Rodrigue's relation (\ref{E8}) and using the properties of Jacobi Polynomial\cite{AbS70, GrR07}, we have
\begin{equation}
y_{n}(z)=N_{n}P_{n}^{(2\delta,-\eta H)}(1 - 2qz).
\label{E25}
\end{equation}
The other part of the wave function obtained from Eq.~(\ref{E9}) as
\begin{equation}
\phi(z) = z^\delta(1 - qz)^{\frac{1}{2}(1  - \eta H)}
\label{E26},
\end{equation}
and therefore, the wave function $\Psi_{n}(z) = \phi (z)y_{n}(z)$ is given as
\begin{eqnarray}
\Psi_{n}(z)=N_{n}z^\delta(1 - qz)^{\frac{1}{2}(1  - \eta H)}P_{n}^{\left(2\delta, - \eta H\right)}(1-2qz)
\label{E27},
\end{eqnarray}
where $N_n$ is the normalization factor to be determined from the normalization condition
\begin{equation}
\int^1_0{| \Psi_n(z)|}^2\frac{dz}{2z\alpha}=1,
\label{E28},
\end{equation}
that is
\begin{equation}
N_{n}^{2} \int^1_0z^{2c}{(1-qz)}^{(1 + 2d)}{\left[ P_{n}^{(2c, 2d)}(z)\right]}^{2} = 2\alpha,
\label{E29}
\end{equation}
where $c =\delta-\frac{1}{2}$ and $d = -\frac{\eta H}{2} $. Using the following forms of Jacobi polynomial \cite{AbS70, GrR07}
\begin{equation}
P_{n}^{(c, d)}(z) = 2^{-n} \sum^n_{r = 0}(-1)^n \left( \begin{matrix}
   n+c\\
   p  \end{matrix} \right)
\left(\begin{matrix}
   n+d\\
   n-p  \end{matrix} \right){(1-z)}^{n-p}(1-z)^p ,
\label{E30}
\end{equation}
and
\begin{equation}
P_{n}^{( c,d)}(z) = \frac{\Gamma( n+ c + 1 )}{n!\Gamma (n + c + d + 1)}\sum^n_{r=0}\left( \begin{matrix}
   n \\
   r   \end{matrix} \right)\frac{\Gamma(n + c + d + r + 1)}{r + c + 1}{{\left( \frac{z - 1}{2} \right)}^{r}}
\label{E31},
\end{equation}
where $\left( \begin{matrix}
   n\\
   r  \end{matrix} \right) = \frac{n!}{r!\left( n-r \right)!}=\frac{\Gamma(n+1)}{\Gamma\left( r+1\right)\Gamma {(n-r+1)}}$. Now $\left[P_{n}^{(2c,2d)}(x) \right]^2$ can be obtained from Eqs.~ (\ref{E30}) and (\ref{E31}) as
\begin{equation}
\left[ P_{n}^{(2c, 2d)}(z) \right]^{2} = z^{n + r - p}(1 - qz)^{p}A_{nq}(p, r).
\label{E32}
\end{equation}
Substituting Eq.~(\ref{E32}) into Eq.~(\ref{E29}), we have
\begin{equation}
 N_{n}^{2} A_{nq} (p, r) \int^{1}_{0} z^{n + r - p + 2c}(1 - qz)^{1 + 2d + p}dz = 2\alpha
 \label{E33},
\end{equation}
Note that
\begin{equation}
\int^{1}_{0} z^{c - a - 1}(1 - qz)^{-b}dz = {_2F_1 (a, b; a + 1; q)}~\frac{\Gamma(a)\Gamma(c-a)}{\Gamma(c)}~; \label{E34}
\end{equation}
provides $Re(c)<Re(a)>0,| \arg(1 - q)|<\pi$, and
\begin{equation}
\int^1_0 z^{a - 1}(1 - qz)^{-b}dz = \frac{1}{a}[{_2F_1(a,b;a+1;q)}]
\label{E35}.
\end{equation}
Using Eq.~(\ref{E35}), we obtain the normalization factor $N_n$ as
\begin{equation}
N_{n}=\sqrt{\frac{2\alpha B_{nq}(p,r)}{A_{nq}(p,r)}},
\label{E36}
\end{equation}
where
\begin{eqnarray}
&\displaystyle{A_{nq}(p,r)=\frac{(-1)^{n}[\Gamma (n + 2c + 1)]^{2}\Gamma{(n + 2d + 1)}}{\Gamma(n + 2c - p + 1)\Gamma(r + 2d + 1 )\Gamma(2c + 2d + 1)}} \nonumber \\ & \times
\displaystyle{ \sum_{p = 0}^n\frac{(-1)^{p+r}q^{n - p + r}\Gamma{(n + 2c + 2d + r + 1)}}{p! r!(n - p )!(n - r )! \Gamma{(p + 2d + 1)}}},
 \label{E37}
\end{eqnarray}
\begin{equation}
[B_{nq}(p,r)]^{-1} = \frac{1}{n + r - p + 2 \delta + 1}~_{2}F_{1}(n + r - p + 2\delta + 1,\eta H - 1 - p; n + r - p + 2\delta +1; q),
\label{E38}
\end{equation}
\subsection{PT-Symmetry and non-Hermiticity}
When a Hamiltonian commutes with the parity and time reversal operators $PT$, it is called PT-symmetric Hamiltonian, i. e. $[H,~PT] = 0$. The $PT$ operator satisfies the following relations:
\begin{equation}
PpP^{-1} = -p = Tp{{T}^{-1}},~ PxP^{-1} = -x,~ TAT^{-1} = -iA
\label{E40}.
\end{equation}
In this case, the potentials parameter in Eq.~(\ref{E12}) are set as $V_{1},~ V_{2},~V_{3},~ V_{4},~V_{5},~ V_{6},~ q \in {\bf R} ~\mbox{and}~ \alpha \in {\bf I R}  (\alpha \rightarrow i\alpha)$ (the set of purely real numbers and purely complex numbers, respectively), this replacement and  some manipulations in Eq.~(\ref{E12}) lead to the following potential
\begin{equation}
 V^{PT}(x)=\frac{\left\{\begin{matrix}
-V_1\left[1+q(\cos 2\alpha x-i\sin2\alpha x) \right]+V_2\left[ \cos 2\alpha x-isin2\alpha x \right]-4(V_3+V_6)\\
-V_4\left[( 1-q^2)\cos 2\alpha x+i(1+q^2)\sin 2\alpha x \right]+V_5\left[( 1+q^2)\cos 2\alpha x+i(1-q^2)\sin 2\alpha x-2q \right]\end{matrix} \right\}}{\left( 1+{{q}^{2}} \right)\cos 2\alpha x+\left( 1-{{q}^{2}} \right)isin2\alpha x+2q}
 \label{E41}.
\end{equation}
Using Eqs~(\ref{E11}) and (\ref{E41}) and making the corresponding parameter replacements in Eqs. (\ref{E24}) and (\ref{E28}), we obtain the energy eigenvalue equations and eigenfunctions, respectively as
\begin{eqnarray}
&&E^{PT}_{n}= \frac{\alpha^{2}}{4} \left[ \frac{\frac{V_{1} - 2q V_{4} - q V_{5}}{q \alpha ^{2}} - \frac{V_{2}}{q^{2} \alpha ^{2}} }{\left(2n + 1 - \eta H^{PT}\right)}-\left( 2n + 1 - \eta H^{PT}\right)\right]- V_{4}, ~  n\ge 0,q\ge 1\\ 
&&\mbox{with} \ \ H^{PT}=\sqrt{1 - \frac{V_{2}}{q^{2}\alpha^{2}} - 4\frac{(V_{3} + V_{6})}{q \alpha ^{2}} - 3\frac{V_{5}}{\alpha^{2}}}\nonumber
\label{E42}
\end{eqnarray}
and
\begin{eqnarray}
\Psi^{PT}_{n}(z) &=&C_{n}z^{\delta}(1 - qz)^{\frac{1}{2}\left(1 - \eta H^{PT} \right)}P_{n}^{\left(2\delta, -\eta H^{PT} \right)}(1 - 2qz) .
\label{E43}
\end{eqnarray}
Using  Eq.(\ref{E43}), we can consequently obtain the normalization constant $C_n$ as
\begin{equation}
C_n=\sqrt{\frac {2\alpha B^{PT}_{nq}(p,r)}{{{A}^{PT}_{nq}}( p,r)}},
\label{E44}
\end{equation}
with
{\small
\begin{eqnarray}
A^{PT}_{nq}(p,r)&=&\frac{{(-1)^{n}}{{\left[\Gamma\left(n+2c+1 \right) \right]}^{2}}\Gamma{(n+2d^{PT}+1)}}{\Gamma\left( n+2c-p+1 \right)\Gamma\left( r+2d^{PT}+1 \right)\Gamma(2c+2d^{PT}+1)} \nonumber \\
&\times&\sum_{p = 0}^n\frac{{\left(-1\right)}^{p+r}{{q}^{n-p+r}}\Gamma{(n+2c+2d^{PT}+r+1)}}{p!r!\left(n-p \right)!\left(n-r \right)!\Gamma{(p+2d^{PT}+1)}}
\label{E45}
\end{eqnarray}}
and
\begin{equation}
[B^{PT}_{nq}( p, r)]^{-1}=\frac{1}{n+r-p+2\delta +1}{_2F_1\left( n+r-p+2\delta +1,\eta H^{PT}-1-p;n+r-p+2\delta +1;q \right)}
 \label{E46},
\end{equation}
where
\[d^{PT}=-\frac{\eta H^{PT}}{2},H^{PT}=\sqrt{1-\frac{V_2}{q^2\alpha ^2}-4\frac{V_3+V_6}{q\alpha^2}-3\frac{V_5}{\alpha^2}}.\]
\subsection{Non PT -symmetry and non Hermiticity}
Another form of the potential is obtained by assuming some of the parameters as purely imaginary, i. e. $q\to iq$, $\alpha \to i\alpha$, $V_1\to iV_1$, $V_3\to iV_3$, $V_6\to iV_6$. Under this replacement and after some manipulations, the potential takes the form
{\small
\begin{equation}
V^{nPT}(x) = \frac{\left\{\begin{matrix}
   V_{1} [ q \cos{\alpha x } + i(1 + q \cos{2\alpha x})] + V_{2}[\cos{2\alpha x} - i\sin{2\alpha x }] - i(V_{3} + V_{6})  \cr - V_{4} [(1 + q^{2}) \cos{ 2\alpha x} + i(1 - q^{2}) \sin{2\alpha x}] + V_{5}[(1 + q^{2})\cos{2\alpha x} + i( 1-q^{2} ) \sin {2\alpha x} - 2iq ]\end{matrix} \right\}} {( 1 - q^{2})\cos{ 2\alpha x} + i(1 + q^{2})\sin{ 2\alpha x} + 2iq}
\label{E47}.
\end{equation}}
Similarly, we obtain the energy eigenvalue equation and eigenfunction, respectively as
\begin{eqnarray}
E^{nPT}_{n} &=& \frac{\alpha^{2}}{4}\left[ \frac{\frac{{{V}_{1}}-2q{{V}_{4}}-q{{V}_{5}}}{q{{\alpha }^{2}}}-\frac{{{V}_{2}}}{{{q}^{2}}{{\alpha }^{2}}}}{2n+1-\eta H^{nPT}}-\left( 2n+1-\eta H^{nPT} \right) \right]^{2}-V_4;\ \ \ n\ge0, q\ge 1\\
&&\mbox{with}\ \ H^{nPT}=\sqrt{1+\frac{{{V}_{2}}}{{{q}^{2}}{{\alpha }^{2}}}-4\frac{({{V}_{3}}+{{V}_{6}})}{q{{\alpha }^{2}}}-3\frac{{{V}_{5}}}{{{\alpha }^{2}}}}
\label{E48}
\end{eqnarray}
and
\begin{eqnarray}
{{\Psi }^{nPT}_{n}}\left( z\right)&=&{{D}_{n}}{z}^{\delta}{{\left( 1-qz \right)}^{\frac{1}{2}\left( 1-\eta H^{nPT}\right)}}P_{n}^{\left( 2\delta,-\eta H^{nPT} \right)}(1-2qz)
\label{E49},
\end{eqnarray}
where the normalization constant $D_n$ is obtain as
\begin{equation}
{{D}_{n}}=\sqrt{\frac{2\alpha{{B}^{nPT}_{nq}}(p,r)}{{{A}^{nPT}_{nq}}\left( p,r \right)}}
\label{E50},
\end{equation}
and
{\small
\begin{eqnarray}
A^{nPT}_{nq}(p,r)&=&\frac{{(-1)^{n}}{{\left[\Gamma\left(n+2c+1 \right) \right]}^{2}}\Gamma{(n+2d^{nPT}+1)}}{\Gamma\left( n+2c-p+1 \right)\Gamma\left( r+2d^{nPT}+1 \right)\Gamma(2c+2d^{nPT}+1)}\nonumber\\
&\times& \sum_{p = 0}^n\frac{{\left(-1\right)}^{p+r}{{q}^{n-p+r}}\Gamma{(n+2c+2d+r+1)}}{p!r!\left(n-p \right)!\left(n-r \right)!\Gamma{(p+2d+1)}}
 \label{E51},
\end{eqnarray}}
\begin{equation}
\left[B^{nPT}_{nq}(p,r)\right]^{-1}=\frac{1}{n+r-p+2\delta +1}{_2F_1\left( n+r-p+2\delta +1,\eta H^{nPT}-1-p;n+r-p+2\delta +1;q \right)}
\label{E52}
\end{equation}
\[d = -\frac{\eta H^{nPT}}{2},~ H^{nPT}=\sqrt{1+\frac{V_2}{q^2\alpha ^2}-4\frac{(V_3+V_6)}{q\alpha^2}-3\frac{V_5}{\alpha^2}}.\]
\section{Special cases}
In this section, in the framework of the NU method, the energy eigenvalues and the corresponding wave functions for the Schr\"{o}dinger equation with the $q$-deformed Woods--Saxon potential, Rosen--Morse potential and symmetrical double well potential are obtained by setting appropriate parameters in Eq.~(\ref{E12}) to zero. The PT-symmetric and the non PT-symmetric solution of this potential are also considered.
\subsection{Woods--Saxon potential}
If $V_3 = V_4 = V_5 = V_6 = 0$, the potential in Eq.~(\ref{E12}) turns to the Woods--Saxon  potential [1 - 5] 
\begin{equation}
V^{{\rm WS,1}}(x)=\frac{-V_{1} {\rm e}^{-2\alpha x}}{1 + q {\rm e}^{-2\alpha x}} + \frac{- V_{2}{{\rm e}^{-4\alpha x}}}{1 + q{\rm e}^{-2\alpha x}}
\label{E53},
\end{equation}
the energy-eigenvalue equation and wave function are obtained, respectively as
\begin{eqnarray}
E^{{\rm WS},1}_{n} &=& -\frac{\alpha^2}{4} \left[\frac{\frac{V_1}{q\alpha^2}+\frac{V_2}{q^2\alpha ^2}}{\left(2n+1-H^{{\rm WS},1}\right)}-(2n+1-H^{{\rm WS},1})\right]^2,~~ n\ge0,~ q\ge1,~~ \eta=1
\label{E54}
\end{eqnarray}
and
\begin{eqnarray}
\Psi^{{\rm WS},1}_{n}(z) &=& {\rm P}_{n}z^{\sqrt{\xi}}(1 - qz)^{\frac{1}{2}\left(1 - \sqrt{H^{{\rm WS},1}}\right)}{\rm P}_{n}^{\left(2 \sqrt{\xi},~-\sqrt{H^{{\rm WS},1}}~ \right)}(1 - 2qz).
\label{E55}
\end{eqnarray}
where \[ c^{\rm WS}= \sqrt{\xi}-\frac{1}{2},~ d^{\rm WS} = -\frac{ H^{{\rm WS},1}}{2},~~ H^{{\rm WS},1}=\sqrt{\frac{V_2}{q^2\alpha^2}+1}.\]
These results are consistent with the results of Meyur and Debnath (2010) \cite{MeD10}. Using Eq.~(\ref{E55}), the normalization constant $P_n$ is obtained as
\begin{equation}
P_{n}=\sqrt{\frac{2\alpha B_{nq}(p,r)}{A_{nq}(p,r)}}
\label{E56},
\end{equation}
where
{\small
\begin{eqnarray}
A^{\rm WS}_{nq}(p,r) &=& \frac{{(-1)^{n}[\Gamma(n + 2c^{\rm WS} + 1)]}^{2} \Gamma{(n+2d^{\rm WS}+1)}}{\Gamma(n + 2c^{\rm WS} - p + 1)\Gamma(r + 2d^{\rm WS} + 1 ) \Gamma(2c + 2d^{\rm WS} + 1)}\nonumber\\
&\times&\sum_{p = 0}^n\frac{(-1)^{p + r}q^{n - p + r} \Gamma{(n + 2c^{\rm WS} + 2d^{\rm WS} + r + 1)}}{p! r! (n - p)! (n - r)! \Gamma{(p + 2d^{\rm WS} + 1)}}
 \label{E57},
\end{eqnarray}}
\begin{equation}
\left[ B^{\rm WS}_{nq}(p,r) \right]^{ -1} = \frac{1}{n + r - p + 2 \delta + 1}~ {_2F_1 ( n + r - p + 2 \delta + 1,\eta H^{{\rm WS},1} - 1 - p; n + r - p + 2\delta + 1; q) }
 \label{E58},
\end{equation}

Now, the case in which $\alpha$ is a pure imaginary parameter is considered, i.e., $\alpha \to i\alpha$, after this substitution and some manipulations, the potential takes the form
\begin{equation}
V^{WS,2}(x) = \frac{-V_{1}[1 + q(\cos{2\alpha x} - i\sin{2\alpha x})] + V_{2}[\cos{2\alpha x} - i\sin{2\alpha x}]}{( 1 + q^{2} ) \cos{ 2\alpha x} + (1 - q^{2}) i\sin{2\alpha x} + 2q}
\label{E59}.
\end{equation}
The real positive energy eigenvalues are obtained as
\begin{eqnarray}
E^{WS,2}_n&=&\frac{\alpha^2}{4}\left[ \frac{\frac{V_1}{q\alpha^2}-\frac{V_2}{q^2\alpha^2}}{2n+1-\sqrt{H^{{\rm WS},2}}}-\left( 2n+1-\sqrt{H^{{\rm WS},2}} \right) \right]^{2},\ \eta =1,\ \ n\ge 0,\ \ q\ge 1,
\label{E60}
\end{eqnarray}
where
 \[ H^{{\rm WS},2}=\sqrt{1-\frac{V_2}{q^2\alpha^2}}.\]

The above equation (\ref{E60}) is consistent with the results of Meyur and Debnath (2010) \cite{MeD10} and Ikhdair and Sever (2007) \cite{IkS07}  for $\hbar = 2\mu = 1$ and $\alpha =\frac{\alpha_1}{2}$. By assuming some of the parameters as pure imaginary, i. e. $q \to iq$, $\alpha \to i\alpha$ and $V_1\to iV_1$, then, we have potential of the form
\begin{equation}
V^{WS,3}(x) = \frac{V_{1}[ q\cos{\alpha x} + i (1 + q\cos{2\alpha x})] + V_{2}[\cos{2\alpha x} - i\sin{2\alpha x }]}{( 1 - q^{2}) \cos{ 2\alpha x} + i (1 + q^{2})\sin{ 2\alpha x} + 2iq}
\label{E61}
\end{equation}
and the corresponding energy eigenvalues is obtain as
\begin{eqnarray}
E^{WS,3}_{n} &=&\frac{\alpha^{2}}{4}\left[ \frac{\frac{V_{1}}{q\alpha^{2}} - \frac{V_{2}}{q^{2} \alpha ^{2}}}{2n + 1 - \sqrt{H^{{\rm WS},1}}}- \left(2n + 1 - \sqrt{H^{{\rm WS},1}} ~\right) \right]^{2},~~ n\ge 0,~~ q\ge 1.
\label{E62}
\end{eqnarray}
\subsection{Rosen--Morse Potential}
If $V_1=V_2  = V_5 =  V_6  = 0$, then the potential in Eq.~(\ref{E12}) reduces to the Rosen--Morse  potential \cite{Oye12, RoM32+, QE1, QE2, QE3, QE4}
\begin{equation}
V^{\rm RM,1}(x) = -V_{3}sech_{q}^{2}(\alpha x) - V_{4}\tanh_{q}(\alpha x).
\label{E63}
\end{equation}
The energy eigenvalue equation and corresponding wave function are obtain respectively as:
\begin{eqnarray}
E^{\rm RM,1}_{n} = -\frac{\alpha^{2}}{4}\left(2n + 1 + H^{RM,1}\right)^{2}- \frac{V_{4}^2}{\alpha^2}\left(2n + 1 +H^{RM,1}\right)^{-2},~~ n\ge 0,~~ q\ge 1
\label{E64}
\end{eqnarray}
and
\begin{eqnarray}
\Psi_n(z) = R_{n}z^{\delta}(1-qz)^{\frac{1}{2}\left(1+H^{RM,1}\right)}P_{n}^{\left( 2\delta, H^{RM,1}\right)}(1-2qz),\ \eta =-1
\label{E65}.
\end{eqnarray}
These results are consistent with the results of Akbarieh and Motavalli (2008) \cite{AkM08}, Ikhdair (2010) \cite{Ikh10} and Oyewumi (2012) \cite{Oye12} for the relativistic limits.
Using the above equation (\ref{E65}), we obtained the normalization constant $R_n$ as
\begin{equation}
R_{n}^{RM,1} = \sqrt{\frac{2\alpha B^{RM,1}_{nq}(p,r)}{A^{RM,1}_{nq}(p,r)}},
\label{E66},
\end{equation}
where
{\small
\begin{eqnarray}
A^{RM,1}_{nq}(p,r)&=&\frac{(-1)^{n}\left[\Gamma(n + 2c + 1) \right]^{2}\Gamma{(n+2d+1)}}{\Gamma( n+2c-p+1)\Gamma( r+2d+1) \Gamma(2c+2d+1)} \nonumber\\
&&\sum_{p = 0}^n\frac{(-1)^{p+r}{q^{n - p + r} \Gamma{(n+2c+2d+r+1)}}}{p!r!(n-p)!(n-r )! \Gamma{(p+2d^{RM,1}+1)}}
 \label{E67},
 \end{eqnarray}}
 \begin{equation}
\left[B^{RM,1}_{nq}(p,r) \right]^{-1} = \frac{1}{n+r-p+2\delta +1}~{_2F_1\left( n+r-p+2\delta +1,\eta H-1-p;n+r-p+2\delta +1;q \right)}
\label{E68},
\end{equation}
\[d = \frac{ H^{RM,1}}{2},~~ H^{RM,1} = \sqrt{\frac{V_{3}}{q\alpha^{2}} + 1}.
\]
Again, a case in which $\alpha$ is a pure imaginary parameter is considered, i.e., $\alpha \to i\alpha$, then equation (\ref{E63}) takes the form
\begin{equation}
 V^{RM,2}(x) = \frac{-4V_{3} - V_{4} \left[(1 - q^{2} )\cos{ 2\alpha x} + i(1 + q^{2})\sin{ 2\alpha x} \right]}{(1 + q^{2} ) \cos{ 2\alpha x} + (1 - q^{2})i\sin{2\alpha x} + 2q}
\label{E69}
\end{equation}
The energy eigenvalues is given as
\begin{eqnarray}
E^{RM,2}_{n} &=& \frac{\alpha^{2}}{4}\left[ \frac{\frac{2V_{4}}{\alpha^{2}}}{\left(2n + 1 - H^{RM,2}\right)}- \left(2n + 1 - H^{RM,2}\right) \right]^{2} - V_{4},~~ n\ge 0,~q\ge 1\nonumber\\
&&\mbox{with}\ \ H^{RM,2}=\sqrt{1 - 4\frac{V_{3}}{q\alpha ^{2}}}
 \label{E70},
\end{eqnarray}
Now, by setting some of the parameters as pure imaginary, i.e., $q\to iq$  ,$\alpha\to i\alpha$, $V_3\to iV_3$ and $V_4\to iV_4$, then equation (\ref{E63}) takes the form
\begin{equation}
 V^{RM,3}(x) = \frac{-iV_{3} - V_{4} \left[(1 + q^{2}) \cos{ 2\alpha x} + i(1 - q^{2}) \sin{ 2\alpha x} \right]}{(1 - q^{2}) \cos{ 2\alpha x} + i(1 + q^{2}) \sin{ 2\alpha x} + 2iq}
  \label{E71}.
\end{equation}
The energy eigenvalue equation is also found to be the same as equation (\ref{E70}).
\subsection{Symmetrical double-well potential}
Choosing $V_1  =V_2  = V_3  =V_4  = 0$ in Eq.~(\ref{E12}), we have
\begin{equation}
V^{SDW,1}(x) = V_{5} \tanh_{q}^{2}{(\alpha x)} - V_{6} sech_{q}^{2}{(\alpha x)}.
\label{E72}
\end{equation}
Then, energy eigenvalues and the corresponding wave functions are obtain respectively as:
\begin{eqnarray}
E^{SDW,1}_{n} &=& -\frac{\alpha^{2}}{4}\left[ \frac{\frac{V_{5}}{\alpha^{2}}}{2n + 1 - \eta H^{SDW,1}}\left(2n + 1 - \eta H^{SDW,1} \right) \right]^{2},~~ n\ge 0,~~ q\ge 1
\label{E73}
\end{eqnarray}
and
\begin{eqnarray}
\Psi_n(z) &=& W_{n}z^{\sqrt{\xi }}(1 - qz)^{\frac{1}{2}\left(1 - \eta H^{SDW,1}\right)}P_{n}^{\left(2\sqrt{\xi},~-\eta H^{SDW,1}\right)}(1 - 2qz),~\eta = \pm 1~.
\label{E74}
\end{eqnarray}
Using the above equation, we obtained the normalization constant $W_n$ as
\begin{equation}
W_n=\sqrt{\frac{2\alpha B^{SDW,1}_{nq}(p,r)}{A^{SDW,1}_{nq}( p,r)}}
\label{E75},
\end{equation}
where
{\small
\begin{eqnarray}
A_{nq}(p,r) &=& \frac{(-1)^{n}[\Gamma (n + 2c^{WS} + 1)]^{2} \Gamma{(n + 2d^{SDW,1} + 1)}}{\Gamma(n + 2c^{WS} - p + 1)\Gamma(r + 2d^{SDW,1} + 1) \Gamma(2c^{WS} + 2d^{SDW,1} + 1)}\nonumber\\
&& \sum_{p = 0}^n \frac{(-1)^{p+r}q^{n-p+r} \Gamma{(n + 2c^{WS} + 2d^{SDW,1} + r + 1)}}
{p!r!(n - p)! (n - r)! \Gamma{(p + 2d + 1)}},
\label{E76}
\end{eqnarray}}
\begin{equation}
\left[B^{SDW,1}_{nq}(p, r)\right]^{-1} = \frac{1}{n + r - p + 2\delta + 1}~{_2F_1(n + r - p + 2\delta + 1, \eta H^{SDW,1} - 1 - p; n + r - p + 2\delta + 1; q )},
\label{E77}
\end{equation}
\[ d^{SDW,1} = -\frac{\eta H^{SDW,1}}{2},~ H^{SDW,1} = \sqrt{4\frac{V_6}{q\alpha^2} + \frac{3V_{5}}{\alpha^{2}} + 1}.\]

Let us consider the case where at least one of the potential parameters is complex. If $\alpha$ is a pure imaginary parameter, i. e. $\alpha\to i\alpha$, symmetrical double well potential can be written as
\begin{equation}
V^{SDW,2}(x) = \frac{-4V_{6} + V_{5}[(1 + q^{2})\cos{ 2\alpha x} + i(1 - q^{2})\sin{ 2\alpha x} - 2q]}{(1 + q^{2})\cos{ 2\alpha x} + (1 - q^{2}) i\sin{2\alpha x} + 2q}.
\label{E78}
\end{equation}
By using Eqs.~ (\ref{E11}) and (\ref{E77}), and making the corresponding parameter substitutions in Eq.~(\ref{E24}), we obtained the real positive energy eigenvalue equation as
\begin{eqnarray}
E^{SDW,2}_{n} &=& \frac{\alpha^{2}}{4}\left[ \frac{\frac{V_{5}}{\alpha^{2}}}{2n + 1 - \eta \sqrt{1 - 4\frac{V_{6} q \alpha^{2}} - 3\frac{V_{5}} \alpha^{2}}}\right.\nonumber\\
&&\left. - \left( 2n + 1 - \eta \sqrt{1 - 4\frac{V_{6}}{ q\alpha^{2}} - 3\frac{V_{5}}{\alpha^{2}}} \right) \right]^{2},~~ n\ge 0,~q\ge 1.
\label{E79}
\end{eqnarray}
When $\alpha \to i\alpha$, $V_6\to iV_6$ and $q \in \Re$ then, $V(x)$ takes the form
\begin{equation}
V^{SDW,3}(x) = \frac{-iV_{6} + V_{5}[(1 + q^{2})\cos{ 2\alpha x} + i(1 - q^{2})\sin{2\alpha x} - 2iq]}{(1 - q^{2}) \cos{ 2\alpha x} + i(1 + q^{2})\sin{ 2\alpha x} + 2iq}
 \label{E80}
\end{equation}
The positive energy eigenvalues is also found to be the same as the one obtained in Eq.~(\ref{E79}).

\section{Conclusion}
In this paper we solve the Schr\"{o}dinger equation for Woods--Saxon potential plus Rosen--Morse potential plus Symmetrical double well potential has been solved via NU method. The results include the energy spectrum of Wood-Saxon potential, Rosen--Morse potential and Symmetrical double well potential  obtained by setting some parameters to zero. The PT-symmetric and non-PT-symmetric of Wood-Saxon potential, Rosen--Morse potential and the symmetrical double well potential are also discussed.

{\bf Acknowledgments}

{ \footnotesize We are grateful to Profs. S. H. Dong, C. Berkdemir, M. Motavalli and O. Aydo$\check{g}$du for communicating some of their research materials to us. Also, KJO acknowledges eJDS (ICTP). }

\end{document}